%Paper: hep-ph/9211293
%From: KRASTEV%UNCVX1.BITNET@VTVM2.CC.VT.EDU
%Date: Fri, 20 Nov 92 19:40 EDT
%Date (revised): Fri, 20 Nov 92 21:07 EDT

%PLAIN TEX file
\magnification 1200

\def\dm{$\Delta m^2$\hskip 0.2cm }

\font\eightrm=cmr8
\line{\hfil CERN--TH.6648 /92}
\line{\hfil IFP--458--UNC}
\vskip 2cm
\centerline{\bf TIME VARIATIONS OF SOLAR NEUTRINO SIGNALS}
\centerline{\bf AND THE RSFCN HYPOTHESIS}

\vskip 1cm
\centerline{P.I. Krastev\footnote*{\eightrm
      Permanent address: Institute of Nuclear Research and
Nuclear Energy, Bulgarian Academy of Sciences,
                   BG--1784 Sofia, Bulgaria; Address after September:
Institute of Field Physics,
Department of Physics and Astronomy, University of North Carolina,
 Chapel Hill, North Carolina 27599--3255, U.S.A.}}

\vskip 1cm

\centerline{Theory Division}

\centerline{CERN}
\centerline{CH--1211 Geneva 23}
\centerline{Switzerland}

\vskip 2.5cm

\centerline{{\bf Abstract}}
\vskip 0.5cm
Resonant spin--flavour conversion of neutrinos (RSFCN) in twisting
magnetic fields might be at the origin of the apparent anticorrelation
between the $^{37}$Ar production--rate in the Homestake solar neutrino
detector and the solar activity. Moreover, it can account for the
results of all solar neutrino experiments reported so far including
the recent results of GALLEX and SAGE.

\vskip 3.1cm
  CERN--TH.6648/92

  September 1992

\footline{\hfil}
\vfill\eject
\count0=1
\footline{\hss\tenrm\folio\hss}
{\bf 1.Introduction}
One of the mysteries of the solar neutrino problem [1] remains the
variations of the $^{37}$Ar production rate in the Cl--Ar solar
neutrino detector [2]. The analyses made so far show that it is
difficult to rule out unambiguously the hypothesis that the $^{37}$Ar
production--rate anticorrelates with the solar activity cycle [3].
If they can be measured accurately, the variations of the signals
in solar neutrino detectors  could be a sensitive test of the
different solutions of the solar neutrino problem. In particular, the
solution in terms of neutrino flavour conversion (MSW effect) allows
us to account for relatively weak variations of the detection rates
which are rather unlikely to be anticorrelated with the solar activity
cycle [4].
  Feeble variations of the neutrino event--rates are expected if
g--mode oscillations of the sun are lowering the temperature in
the centre of the sun, T$_{\rm c}$,  thereby affecting the highly
temperature--dependent $^8$B and $^7$Be neutrinos [5].\footnote*{This
solution of the solar neutrino problem cannot explain the relative
suppressions of the experimentally measured signals as compared with
the predicted  ones
in the standard solar model [1], in particular the fact that the
suppression measured in Kamiokande is weaker than the one
measured in the Cl--Ar experiment. Exactly the opposite effect to
the observed one is predicted in all non--standard solar models
with lower T$_{\rm c}$ (see ref.[6]).}
      Larger variations can be accommodated in the spin--precession
[7] or in the                                     resonant
spin--flavour conversion scenario [8]. However, with several detectors
simultaneously taking data it becomes a challenge to reconcile their
results. In particular, the 20--year data sample of the Cl--Ar
experiment of R. Davis and his collaborators seems to indicate
strong variations by a factor larger than two of the $^{37}$Ar
production--rate in anticorrelation with the solar activity.
On the other hand, the Kamiokande--II collaboration reported results
[9] which are consistent with no or only small variations of the
event rate in their $\nu_e e$--scattering water--Cherenkov
detector. Attempts to reconcile these results have been made already.
However, the problem, if taken seriously, is far from being solved. In
particula
   r,
the solution found in [10] does not reproduce the mean values
of the signals in the Cl--Ar and Kamiokande experiments with one and
the same magnetic field strength,               the discrepancy
being larger than the one--sigma error bars of these experiments.
Moreover, a large magnetic field in the radiation zone is needed in
order for the signal to be suppressed in years of quiet sun. The
solution of ref. [11] generally satisfies the first requirement but the
authors do not discuss at all the second one. An unrealistic magnetic
field distribution is used and strong suppression of the $^{71}$Ge
production--rate is predicted which contradicts the latest results
from the Ga--Ge solar neutrino experiments.

  In the present note a new scenario is proposed reconciling the
variation patterns of the signals in all solar neutrino detectors
taking data at present. It is based on the spin--flavour conversion
of neutrinos in twisting magnetic fields. As first shown in [12]
the dynamics of this effect differs considerably from the case in which
the direction of the transverse magnetic field is fixed in space [8].
The neutrino transitions have a resonant character even in the case
of a Dirac--neutrino mass--spectrum. The resonant condition comprises
dependence on the angular velocity of the rotating magnetic field and
the adiabatic condition  on its second derivative. The possible
consequences of this effect for solar neutrinos have been discussed in
[13]. In particular, it was shown that the suppression factors change
most dramatically for large E/\dm which corresponds to almost degenerate
 or Dirac--neutrino masses. This is actually just the opposite of what
is needed for explaining the simultaneous absence of variations
in the Kamiokande detector and the presence of relatively strong
variations in the Cl--Ar one. It is well  known that in Kamiokande
only high--energy $^8$B--neutrinos can be detected, whereas in Cl--Ar
detectors both $^8$B-- and $^7$Be--neutrinos contribute to the
signal. Therefore, one expects in general that for a fixed magnetic
field distribution the rotation will give a result opposite to the
desired one, namely the signal in Kamiokande will vary more strongly
than in the Cl--Ar detector. However, it seems plausible that
the twist of the magnetic field should be correlated with its
strength. In the absence of a detailed model of the solar magnetic
field one can hardly predict the relation between
the strength of the magnetic field and  its
twist. More twisted fields might be either stronger or weaker
depending on the specific structure of the magnetic field lines.
As shown below, by adjusting the strength of the magnetic field and of
its twist one can reconcile the different variation patterns in these two
 detectors. Moreover, relatively small variations for Ga--Ge detectors
are  naturally obtained.
 \vskip 0.5cm
{\bf 2. Solar neutrino data.}
The mean $^{37}$Ar production rate in the chlorine detector for more
than 20 years of data taking is [14]:
$$Q^{exp}_{Ar} = 2.1 \pm 0.3\hskip 0.2cm {\rm SNU}.\eqno(1)$$
Expressed as a ratio between experimentally measured and theoretically
predicted values of the above quantity in the standard solar model, the
result (1) reads:
$$R_{Ar} = {Q^{exp}_{Ar}\over Q^{th}_{Ar}} = 0.28 \pm 0.03.\eqno(2)$$
In years of maximal solar activity, from 1979.5 to 1980.7 and from 1988.4
  to 1989.5 this ratio drops to
$$R_{Ar}^{min} = 0.075 \pm 0.10,\eqno(3a)$$
$$R_{Ar}^{min} = 0.18 \pm 0.10,\eqno(3b)$$
respectively and in years of minimum solar activity
from           1977.0 to 1978.0
and  from        1986.8 to 1988.3 it rises to
$$R^{max}_{Ar} = 0.47 \pm 0.14,\eqno(4a)$$
$$R^{max}_{Ar} = 0.45 \pm 0.10,\eqno(4b)$$
respectively. Note that the period from  1988.4 to 1989.5
precedes the maximum of the 22nd solar cycle at the end of 1990.
In the Kamiokande--II $\nu$e--scattering
                      experiment during four years, from
1987 until 1990, the ratio between the observed event rate
   and the  predicted one in
the standard solar model is [9]:\footnote*{The latest result of the
Kamiokande--III
      experiment announced at the NEUTRINO '92 conference, namely
$R_{\nu e} = 0.59^{+0.11}_{-0.09} \pm 0.06 $
                during approximately 220 days of data taking
does not differ significantly from the previous ones with the
   same duration                  and changes only
slightly the result eq.(5) (see ref.[15]). Remarkably enough,
however, this might indicate an increase of the signal coinciding with
the descending phase of the solar activity cycle after
                      its maximum in 1990, although a definite
conclusion on this point would be certainly premature and groundless.}
$$R_{\nu e} = {Q^{exp}_{\nu e}\over Q^{th}_{\nu e}} = 0.46 \pm 0.05
\pm 0.06.\eqno(5)$$
During the same period the solar activity has
                 changed from minimum to almost
the maximum of the 22nd solar activity cycle.
                           No significant anticorrelation
between the signal in Kamiokande--II and the solar activity cycle has
been found.   However,
 a small variation within the
error bars cannot be excluded yet.
Therefore in the analysis below the following values of the
mean suppression in years of active and quiet Sun are assumed:
$$R_{\nu e}^{min} = 0.40, \eqno(6a)$$
$$R_{\nu e}^{max} = 0.60. \eqno(6b)$$
The results from the two Ga--Ge experiments SAGE [16] and GALLEX [17]
are:
$$R_{Ge} = 0.52 \pm 0.15,  \eqno(7a)$$
and
$$R_{Ge} = 0.63 \pm 0.21,  \eqno(7b)$$
respectively.
The errors in (7) represent the statistical and
systematic errors added in quadrature.
The data do not  yet allow us to discuss seriously variations of the
signal in this type of detectors. Measurements during at least one full
solar cycle are necessary in order to do that.

The three detectors have different thresholds and therefore they are
sensitive to different parts of the solar neutrino spectrum.
The latter can be roughly divided into three parts. The lowest energy
pp--neutrinos
          with characteristic energy E$_{pp} \approx 0.3$ MeV are
detected only in the Ga--Ge detectors where
they supply slightly more  than half of the signal.
 The intermediate energy
$^7$Be, $^{13}$N, $^{15}$O
                 and pep--neutrinos are detected in Ga--Ge and
Cl--Ar detectors.
                 They  contribute 1.7 SNU to the
signal in Cl--Ar detectors but remain undetected by the Kamiokande
detector because of the high threshold of the latter. The high energy
$^8$B--neutrinos are detected by all three types of detectors.
In fig.1 the percentage of the signal is plotted as a function of the
energy in all three  detectors. Almost 30 \% of the signal in the
Cl--Ar detector comes from a region below the threshold of the
Kamiokande--II detector which is 7.5 MeV during the second half of the
four years of data taking period. During the first two years
the threshold of this detector was 9.3 MeV. The Kamiokande
        collaboration has not observed any change of the event
rate  with respect to the predicted one in the standard solar
model after the improvements that have allowed them to lower the
threshold of their detector.
               For simplicity, in what follows a 7.5 MeV threshold
in this detector is assumed. Irrespective of the threshold, about
14 \% of the $\nu e$--scattering events
               comes from neutral current neutrino--electron
scattering, which is independent of the neutrino flavour.

   Following ref.[18] the relative event rate can be expressed as
sums of products of the partial contributions by the solar neutrinos
that come from different chains of the nuclear reactions in the Sun,
multiplied by the corresponding suppression factors.
 For the mean event rates it follows:
$$R_{Cl} = 0.70 P_H + 0.30 P_I,\eqno(8a)$$
$$R_{\nu e} = 0.14 + 0.86 P_H.\eqno(8b)$$
  From these relations one can easily derive the inequality:
$$R_{Cl} \ge 0.81 R_{\nu e} - 0.11.\eqno(9)$$
This puts a lower limit on the signal in the Cl--Ar experiment for
a given corresponding signal in a $\nu e$--scattering experiment,
e.g. for $R_{\nu e}$ = 0.4 $R_{Cl}$ cannot be smaller than 0.21.
Moreover, one can easily see
  that if the signal in the
  Kamiokande--II detector varies between 0.4 and 0.6 the corresponding
signal in the Cl--Ar experiment should vary between 0.21 and 0.50.

Substituting the mean values from (2) and (5) in (8)
                            one obtains for the relevant
suppression factors:
$${\bar P}_H \simeq 0.37    \eqno(10a)$$
$${\bar P}_I \simeq 0.07    \eqno(10b)$$
For the minimal and maximal event rates measured during years of
active and of quiet Sun respectively, one obtains similarly:
$$P_H^{min} \simeq 0.30  \eqno(11a)$$
$$P_I^{min} \simeq 0.00 \eqno(11b)$$
and
$$ P_H^{max} \simeq 0.53  \eqno(12a)$$
$$P_I^{max}  \simeq 0.43  \eqno(12b)$$

  The problem of reconciliation of the different time--variation
patterns of the signal   in the Cl--Ar detector and in Kamiokande--II
can now be formulated quantitatively. A suitable mechanism has to provide
  the suppression factors (11) and (12) in years of active and of quiet
Sun and, at the same time, the relevant average suppression factors
(10) during an entire solar cycle.
\vskip 1cm
{\bf 3. Time variations in the RSFCN scenario.}
There are different types of neutrino conversion in magnetic
fields. Spin--precession ($\nu_L\rightarrow\nu_R$) is suppressed in
matter and energy independent. It is disfavoured from the experimental
data as the suppressions observed in the chlorine experiment and
in Kamiokande--II e.g., are  significantly different.
Resonant spin--flavour conversion takes place, ($\nu_{eL}\rightarrow
\nu_{\mu R}$ (Dirac) or $\nu_{eL}\rightarrow\bar\nu_{\mu R}$
(Majorana)),                            when the mass splitting
between the different massive neutrino components can be compensated
by the different contributions to the potential energies of each
neutrino component due to their coherent forward scattering in
matter. The Majorana--neutrino case is more favourable for our purpose
as the $\bar\nu_{\mu R}$--neutrinos can contribute to the signal
in $\nu e$--scattering detectors but will remain undetected
in Cl--Ar experiments. Therefore, only for this reason will the signal
in the Cl--Ar experiment vary more strongly than the one in
$\nu e$--scattering experiments. The resonant spin--flavour conversion
allows us to reconcile the mean values (2) and (5) and can in principle
account for the different variation patterns in these experiments [19].
However, considerable fine--tuning of the magnetic field
distribution is necessary in order to get a satisfactory agreement
between the results (3) - (6a) and (4) - (6b), correspondingly (see ref.
[19]).
 Finally, the most promising mechanism capable of producing strong
variations of the signal in the Cl--Ar experiment and much weaker ones
in Kamiokande and in the Ga--Ge ones           seems to be
the neutrino spin--flavour conversion in twisting magnetic fields. The
latter has been discussed in detail in [13] and   reviewed recently
                                        in [20,
21].\footnote*{It was first pointed out in [21] that
the RSFCN in twisting magnetic fields can account for the different
variation patterns in the Cl--Ar and Kamiokande--II experiments.}
The scaling property of the survival probabilities for
   RSFCN in uniformly twisting magnetic fields (see ref.[13] for
details)
                                which leads to a
"stretching" of the suppression curve for E/\dm $>$ (E/\dm)$_{res}$
                                  is the one that
allows us to obtain an almost constant high--energy suppression factor,
$P_H$, a strongly varying intermediate--energy one, $P_I$, and
a close to unity low--energy one, $P_L \approx 1$.

   In fig.2 the probabilities for a left--handed neutrino to remain
a left--handed one are plotted versus E/\dm. They have been obtained by
numerical integration of the system of evolution equations, describing
the transitions between two neutrinos of the ZKM type, e.g.
$\nu_{eL}\leftrightarrow\bar\nu_{\mu R}$. The transition magnetic moment
has been assumed to be $\mu_{\nu}$ $=$ 10$^{-10}$ $\mu_B$.
 The magnetic field
configuration in each  case can be characterized by a pair ($\dot\phi,
\lambda$), where $\dot\phi$ is the angular velocity and $\lambda$ is
a scaling factor describing the change of the magnetic field profile
B$_{\perp}$ $\rightarrow$ $\lambda$B$_{\perp}$ (see fig.3).
The angular velocity, $\dot\phi$, in each case is constant and
the rotation is supposed to take place from the base of the convection
zone (r = 0.7 R$_{\odot}$) to the surface of the Sun (r = R$_{\odot}$).
It is clear already from the behaviour of the suppression factors
in fig.2 that $P_I$ will change strongly, by a factor of 10 at least,
                                          whereas $P_L$ and $P_H$
will vary only within 10 $\div$ 20 \%.
 This is sufficient to reconcile the data from the solar neutrino
experiments under discussion.

For each of the suppression curves in fig.2 the relevant predicted event
rates in the three types of detectors taking data at present were
calculated. Varying \dm one can choose  different suppression
factors, $P_H$, $P_I$ and $P_L$ and correspondingly different
relative event rates, $R_{Ar}$, $R_{\nu e}$ and $R_{Ge}$ can be
obtained.
            The neutrino fluxes and the relevant cross--sections
were taken from ref.[1]. For the Kamiokande--II detector the finite
energy resolution and the trigger efficiency have been taken into
account as described in [22].

In fig.4 the event rates in the Cl--Ar, $\nu e$--scattering and Ga--Ge
experiments are plotted versus $\lambda$. The parameter $\dot\phi$
varies also accordingly but cannot be shown in this two--dimensional
plot. For \dm $=$ $1.12\times 10^{-8}$ eV$^2$ and $\lambda$ $=$ 0.5,
0.85 and 0.92 the values of $R_{Ar}$ and $R_{\nu e}$ are close to
those in eq.(4) and eq.(6b), eq.(3) and eq.(6a) and eq.(2) and
eq.(5), respectively. The corresponding values of $R_{Ge}$ are
0.75, 0.51 and 0.49, which gives for the capture rate in the
Ga--Ge detectors 98.6 SNU, 67.0 SNU and 65.15 SNU respectively. They
are close to those reported by the SAGE and GALLEX collaborations,
eq.(7).
      A good agreement has been achieved between the results from the
different solar neutrino experiments at solar maximum and solar
minimum, as well as for the mean values of the suppressions of the
signals.

        Changing the value of \dm from one half of to two times
the above value destroys this agreement, e.g. for the other two
values in fig.4a and fig.4c the mean values of $R_{Ge}$
$\approx$ 0.31 are one sigma away from the experimentally
measured value eq.(2). Note also that the values of $R_{Ge}$
change quite strongly with \dm. This means that more accurate
data from the Ga--Ge experiments can further  restrict  the
scenario under discussion.

The additional freedom of changing the latitudinal distribution
of both the magnetic field strength and its twist has to be  used
to achieve large annual suppressions, $R_{Ar} < 0.25$,
                                     in years of high
solar activity notwithstanding the semiannual variations due to the
slit near the equatorial plane in the toroidal magnetic field in
the Sun (see ref. [19]).
\vskip 1cm
{\bf Conclusions.}
   The proposed scenario requires significant tuning of the
magnetic field distribution, its overall strength and its twisting
structure, as well as of \dm. However, it shows that it is still
possible to explain both the mean suppression of the signals in all
three detectors taking data at present  and to reconcile the different
variation patterns of the signals in the Cl--Ar and Kamiokande--II
detectors. Quite definite predictions for future solar neutrino
experiments follow from the above considerations and with more data
this solution of the solar neutrino problem will be tested further.
\vskip 1cm
{\bf Note Added.} In a recent preprint [23] the variation of the
signals in the Cl--Ar and  in the Kamiokande--II experiments have been
studied in the twisting toroidal magnetic fields scenario. More detailed
structure of the magnetic field distribution and of its twsiting
structure have been utilized however  no predictions for Ga-Ge experiments
have been obtained and the twenty year mean value of the $^{37}$Ar
production rate for the Cl--Ar
experiment in this paper seems to be in conflict with the measured one.
\vfill\eject
\centerline{\bf References}
\vskip 1cm
\item{1.} J.N. Bahcall and R.K. Ulrich, Rev.Mod.Phys. 60 (1988) 298,
and J.N. Bahcall, Talk given at the Int. Workshop on Electroweak Physics
Beyond the Standard Model, Valencia, Spain, 1991.
\item{2.} R. Davis et al., XXI Int. Cosmic Ray Conf., 6 - 19 January,
Adelaide, Australia, Conference papers, vol.7, p.155, 1990.
\item{3.} G.A. Bazilevskaya, Yu.I. Stozhkov  and T.N. Charakhch'yan,
Pis'ma Zh.ETF, 35 (1982) 273 (Soviet Phys.--JETP Letters, 35 (1982) 341);
J.N. Bahcall and H. Press, Ap. J. 370 (1991) 730; B.W. Filippone and
P. Vogel, Phys. Lett. B246 (1990) 546; J.W. Bieber, D. Seckel,
T. Stanev and G. Steigman, Nature 348 (1990) 407; H. Nunokawa and
H. Minakata, Proc. of the 25th Int. Conf. on High Energy Physics,
Singapore, 1990, p.681.
\item{4.} P.I. Krastev and A.Yu. Smirnov, Mod. Phys. Lett. A6 (1991)
1001.
\item{5.} A. De R\'ujula and S.L. Glashow, preprint HUTP-92/A038 and
CERN--TH 6608/92, unpublished.
\item{6.} P.I. Krastev, S.P. Mikheyev and A.Yu. Smirnov, in Proc. of
the Internationsl School "Low Energy Weak Interactions" (LEWI'90),
Dubna, September 4  - 13 (1990), Editions JINR, Dubna (1991), p.96.
\item{7.} M.B. Voloshin and M.I. Vysotski, Sov. J. Nucl. Phys.
44 (1986) 845;
 L.B. Okun, M.B. Voloshin and M.I. Vysotski, Sov. Phys. JETP
64 (1986) 446.
\item{8.} C.-S. Lim and W.J. Marciano, Phys. Rev. D37 (1988) 1368;
E.Kh. Akhmedov, Phys. Lett. B213 (1988) 64.
\item{9.} K.S. Hirata et al., Phys. Rev. D44 (1991) 2241, and
references therein.
\item{10.} K.S. Babu, R.N. Mohapatra and I.Z. Rothstein, Phys. Rev.
D44 (1991) 2265. The scenario studied in this paper has been briefly
mentioned previously by E. Akhmedov in Phys. Lett. B257 (1991) 163.
\item{11.} Y. Ono and D. Suematsu, Phys. Lett. B271 (1991) 165.
\item{12.} A.Yu. Smirnov, Pis'ma ZhETF 53 (1991) 280; Phys. Lett.
B260 (1991) 161.
\item{13.} E.Kh. Akhmedov, P.I. Krastev and A.Yu. Smirnov, Z. Phys.
C - Particles and Fields, 52 (1991) 701.
\item{14.} R. Davis et al., XXI Int. Cosmic Ray Conf., January 6--19,
Adelaide, Australia, Conference papers, vol. 7, p. 155, 1990.
\item{15.} K. Nakamura, talk given at the NEUTRINO '92 conference,
7 - 12 June 1992, Grenada, Spain, (to be published in the proceedings).
\item{16.}
  V. Gavrin, Talk given at the XXVIth  Int. Conference
on High Energy Physics, August 5 - 13, 1992, Dallas (U.S.A.).
\item{17.} P. Anselmann et al., (GALLEX collaboration), Phys. Lett.
B285 (1992)
\item{18.} V. Barger, R.J.N. Phillips and K. Whisnant,  Phys. Rev.
D43 (1991) 1110.
\item{19.} P.I. Krastev and A.Yu. Smirnov, Z. Phys. C - Particles and
Fields, 49 (1991) 675; R. Barbieri and G. Fiorentini, in Second
International Workshop on Neutrino Telescopes, Venezia, 1990,
ed. by Milla Baldo Ceolin, p.47.
\item{20.} E.Kh. Akhmedov, talk given at the XIIth  Moriond workshop,
"Massive neutrinos. Tests of fundamental symmetries.", Jan. 25 -
Feb.1, 1992,
(to be published in the proceedings).
\item{21.} P.I. Krastev, talk given at the XIIth  Moriond workshop,
"Massive neutrinos. Tests of fundamental symmetries.", Jan. 25 -
Feb.1, 1992,
(to be published in the proceedings).
\item{22.} J. Bahcall and W. Haxton, Phys. Rev. D40 (1989) 931.
\item{23.} T. Kubota, T. Kurimoto, M. Ogura and E. Takasugi, preprint,
Osaka University, OS GE 23 - 92, 1992.

\vfill\eject

\centerline{\bf Figure captions}
\vskip 1.0cm
\item{Fig.1} Percentage of the signal in Ga--Ge (dotted line),
                            Cl--Ar (full line) and    in
$\nu_ e$--scattering (dashed line) experiments as a function
of the energy of the neutrinos.
\vskip 0.5cm
\item{Fig.2} Suppression factors, $P(\nu_L\rightarrow\nu_L)$, as
function of E/\dm for different ($\dot\phi$, $\lambda$) values
            (figures at the curves). The angular velocity
$\dot\phi$ is in units of rad/R$_{\odot}$, where R$_{\odot}$ is
the radius of the Sun.
\vskip 0.5cm
\item{Fig.3} Assumed magnetic field distribution inside the Sun.
 The two curves correspond to $\lambda = 1$ (full line) and
$\lambda = 0.5$ (dashed line) (see text).
\vskip 0.5cm
\item{Fig.4} Suppressions of the signals in Ga--Ge (dot--dashed line)
          Cl--Ar (dashed line) and
Kamiokande (full line)
           experiments as functions of $\lambda$ for different
\dm: a) \dm = 1.12$\times$ $10^{-8}$ eV$^2$,
             b) \dm = 6.31$\times$ $10^{-9}$ eV$^2$ and c) \dm =
3.51$\times$ $10^{-9}$ eV$^2$.

\bye